\pdfoutput=1
\documentclass[10pt,journal,compsoc]{IEEEtran}
\usepackage[normalem]{ulem}
\usepackage{tcolorbox}

\usepackage{graphicx}
\usepackage{booktabs}
\usepackage{pifont}
\usepackage{cite}
\usepackage{amsmath,amssymb,amsfonts}
\usepackage{algorithmic}
\usepackage{graphicx}
\usepackage{textcomp}
\usepackage{comment}
\usepackage{multirow}
\usepackage[numbers, square, comma, sort&compress]{natbib}
\usepackage{caption}
\usepackage{setspace}
\usepackage{csquotes}
\usepackage{hyperref}
\usepackage{diagbox}
\usepackage{spreadtab} 
\usepackage{tablefootnote} 
\usepackage{multirow} 
\usepackage{amsmath} 
\usepackage{mathtools} 
\usepackage{footnote} 
\usepackage{threeparttable}  
\usepackage{enumitem}

\newcounter{reviewer}
\newcounter{point}[reviewer]
\ifCLASSOPTIONcompsoc

\ifCLASSINFOpdf

\else

\fi

\begin{document}
	%
	
	\title{Comprehensive Evaluation of RSB and Spectre Vulnerability on Modern Processors}

	\author{Farhad Taheri,
		Siavash~Bayat-Sarmadi,~\IEEEmembership{Member,~IEEE,}
		Alireza~Sadeghpour,
		Seyed~Parsa~Tayefeh~Morsal
		
		\IEEEcompsocitemizethanks{\IEEEcompsocthanksitem Farhad Taheri~Ardakani, Siavosh Bayat Sarmadi, Alireza Sadeghpour, and Seyed Parsa Tayefeh Morsal are with the Department of Computer Engineering, Sharif University of Technology, Iran. E-Mail: f.taheri89@sharif.edu, sbayat@sharif.edu, sadeghpour@ce.sharif.edu, morsal@ce.sharif.edu
	}}
	
	
	\IEEEtitleabstractindextext{%
		\begin{abstract}
			Performance-enhancing mechanisms such as branch prediction, out-of-order execution, and return stack buffer (RSB) have been widely employed in today’s modern processing units. 
			Although successful in increasing the CPU performance, exploiting the design flaws and security bugs in these components have set the background for various types of microarchitectural attacks such as Spectre and Meltdown.
			While many attacks such as Meltdown and Spectre have been numerously implemented and analyzed on Intel processors, few researches have been carried out to evaluate their impact on ARM processors.
			Moreover, SpectreRSB vulnerability, the newer variant of spectre attack based on RSB, has been neglected in recent studies.
			
			In this work, we first evaluate the SpectreRSB vulnerability by implementing this attack on ARM processors, which, to the best of our knowledge, has not been implemented and analyzed on ARM processors. We further present a security evaluation of ARM processors by implementing different variants of Spectre-family attacks. By analyzing the results obtained from various experiments, we evaluate ARM processors security regarding their diverse microarchitectural designs.
			We also introduce a high throughput and noise-free covert channel, based on the RSB structure. Based on our experiments, the throughput of the covert channel is 94.19KB/s with negligible error. 
		\end{abstract}
		
		}

	\maketitle

	\IEEEdisplaynontitleabstractindextext
	
	\IEEEpeerreviewmaketitle

	\ifCLASSOPTIONcompsoc
	\IEEEraisesectionheading{\section{Introduction}\label{sec:introduction}}
	\else
	\section{Introduction}
	\label{sec:intro}
	\fi
	Since the dawn of modern processors, increasing the performance has long been the main focus of many researches, both in academia and industry field. To achieve the highest performance and maximum efficiency, mechanisms such as cache, branch prediction, out-of-order execution, and return stack buffer (RSB) have been widely employed~\cite{hennessy:13}.
	RSB is widely used to reduce the retrieve time of return address, \cite{RSB:1} whereas out-of-order execution and branch prediction have been utilized to avoid the CPU from running idle, while waiting for required dependencies. This allows the processor to run ahead and execute instructions regardless of their sequential order~\cite{ branch:14, hennessy:13}.
	
	Although such mechanisms have been widely successful to guarantee faster computing, various attacks to compromise the integrity, confidentiality, and availability in a system have been proposed~\cite{c5:26}. 
	Cache, in particular, has been a fundamental playground for various types of attacks. In 2005, Bernstein \cite{c3:24} successfully extracted AES cryptographic key, by the so-called cache attacks.
	Similar attacks were also proposed by Osvik \textit{et al}. \cite{c1:22}, in the following year, obtaining secret data by malicious utilization of cache. The Prime+Probe attack has also been a significant milestone in making cache attacks even more practical. Using the approaches introduced by Prime+Probe and Flush+Reload attacks, cache attacks have surpassed cross-core boundaries. They have been implemented throughout various layers, such as virtual machines (VMs) residing on different physical cores or even against secure architectures such as SGX, web browsers sandboxes, well-known and widely incorporated cryptographic schemes~\cite{c9:30, llcA:7, FR:8}.
	
	In recent years, on top of the old-fashioned cache attacks, a new family of attacks has gained the attention of security experts and CPU manufacturers. Exploiting security bugs in the processors microarchitectural design, named microarchitectural attacks, have exploited out-of-order execution and branch prediction to dismantle a System’s security mechanisms \cite{ Melt:4, Spec:5, m2:35, m3:36, m4:37, m6:39, RSB:1, canella2019fallout}.
	The Meltdown attack \cite{Melt:4}, exploiting out-of-order execution, allows a malicious process to access arbitrary memory in kernel space. The Spectre \cite{Spec:5} attack, uses the branch prediction to break down the isolation between processes. Other attacks have also been proposed, threatening the trusted execution environments (TEE), both on Intel SGX, and \mbox{ARM TrustZone}~\cite{van2018foreshadow:21, m7:40}. 
	These attacks, targeting well-known processors, have posed significant threats to user's security, and mitigating them has become a concern for security researchers, as well as CPU manufacturers such as Intel, ARM, and AMD. The SpectreRSB attack is the recent variant of Spectre-family attack which, exploits RSB to access user sensitive information.~\cite{RSB:1, ret:2}.  
	
	In recent years, with exponential growth in the number of smartphone and internet of things (IoT) devices~\cite{iot3:50}, the security analysis of the ARM processors used in these devices have become an everlasting concern for security specialists \cite{iot6:56}.
	Moreover, with Intel processors being in the highest demand for desktop PCs, servers, cloud computing, and network infrastructures, up-to-date evaluation of their security is utterly critical to maintaining the reliability of computer systems. Furthermore, few researches investigate the vulnerability of RSB comprehensively, e.g., implementing SpectreRSB on ARM processors and implementing a covert channel by RSB. Despite in previous works, many countermeasures and detections method have been proposed~\cite{mazaheri2020lurking, depoix2018detecting, rokicki2020ghostbusters, guarnieri2020spectector, fustos2019spectreguard}; these attacks are still a serious threat to the security of the users.
	
	In this work, we implement the SpectreRSB attack on the ARM processor which, to the best of our knowledge, SpectreRSB has not been implemented on the ARM processor. We demonstrate the importance of these attacks and further investigate effective mitigations to prevent them. 
	Furthermore, we introduce a high throughput noise-free covert channel on Intel processors with X86 architecture through exploiting RSB.
	We demonstrate the effectiveness of our covert channel by implementing the cover channel on high-demand Intel core models and providing the throughput and noise for our real-world attack scenario. 
	We also present a side-channel evaluation of different Intel and ARM processors by implementing and analyzing various types of Spectre attacks, e.g., variant 1, 3/3a, and 4. 
	We provide a security analysis of a wide range of ARM processors that are currently deployed in popular smart phones and IoT devices. These attacks and their impacts on ARM processors have only been investigated in few researches and are limited to some specific ARM core models. Furthermore, we explore fundamental architectural and microarchitectural factors in different models of ARM processors, playing an important role in the possibility of microarchitectural attacks~\cite{arm2:53}.

	\subsection{Our Contributions}
	
	The main contributions of this work are as follows:
	
	\begin{itemize}
		\setlength{\itemsep}{5pt}
		\item Implementing the SpectreRSB attack on ARM processors, which to the best of our knowledge, has not been discussed by any other work.
		
		\item Implementing and investigating Spectre-family attacks (variants 1, 3/3a, and 4) on a range of ARM processors (including Cortex-A8, Cortex-A53, Cortex-A9 and Cortex-A72) concerning their various microarchitectural designs. Based on these processors, our experiments performs on different microarchitectures.
		
		\item Introducing and investigating critical building blocks to implement Spectre-family attacks on different ARM processors, as well as  designing various test scenarios to make a better evaluation of their security.

		\item Introducing a noise-free and high throughput covert channel based on the design flaws in RSB. We evaluate our covert channel on Intel processor and achieve throughput of 94.19KB/s.
		
		
	\end{itemize}
	
	In Section \ref{sec:background}, we review the fundamental concepts of the modern processors. Furthermore, in Section \ref{sec:attacks}, we go over some well-know microarchitectural attacks, which have been either used or re-implemented in this work. 
	In \mbox{Section \ref{sec:building blocks}}, we introduce our essential building blocks for implementing side-channel attacks. Section \ref{sec:evaluation} shows our results and analysis from implementing various attack scenarios and threat models on commonly ARM and Intel processors. In Section \ref{sec:covert channel}, we present our new covert channel based on RSB. In Section \ref{sec:countermeasures}, we investigate various countermeasures that can effectively mitigate Spectre-family attacks. Finally, in Section \ref{sec:conclustion}, we provide our conclusion about various aspects of this work.
	
	\section{Background}
	\label{sec:background}
	In this section, we describe fundamental and required concepts for this paper.
	\subsection{CPU Cache}
	\label{sebsec:cache}
	In today's modern processors, in order to reduce the speed gap between fast CPUs and rather time-consuming memory accesses, memory is managed hierarchically. The more expensive and faster memory units are employed for most recently accessed data, in contrast with cheaper and relatively slower storage units~\cite{hennessy:13}.
	
\begin{figure}[t]
	\centerline{\includegraphics[width=1.05\columnwidth]{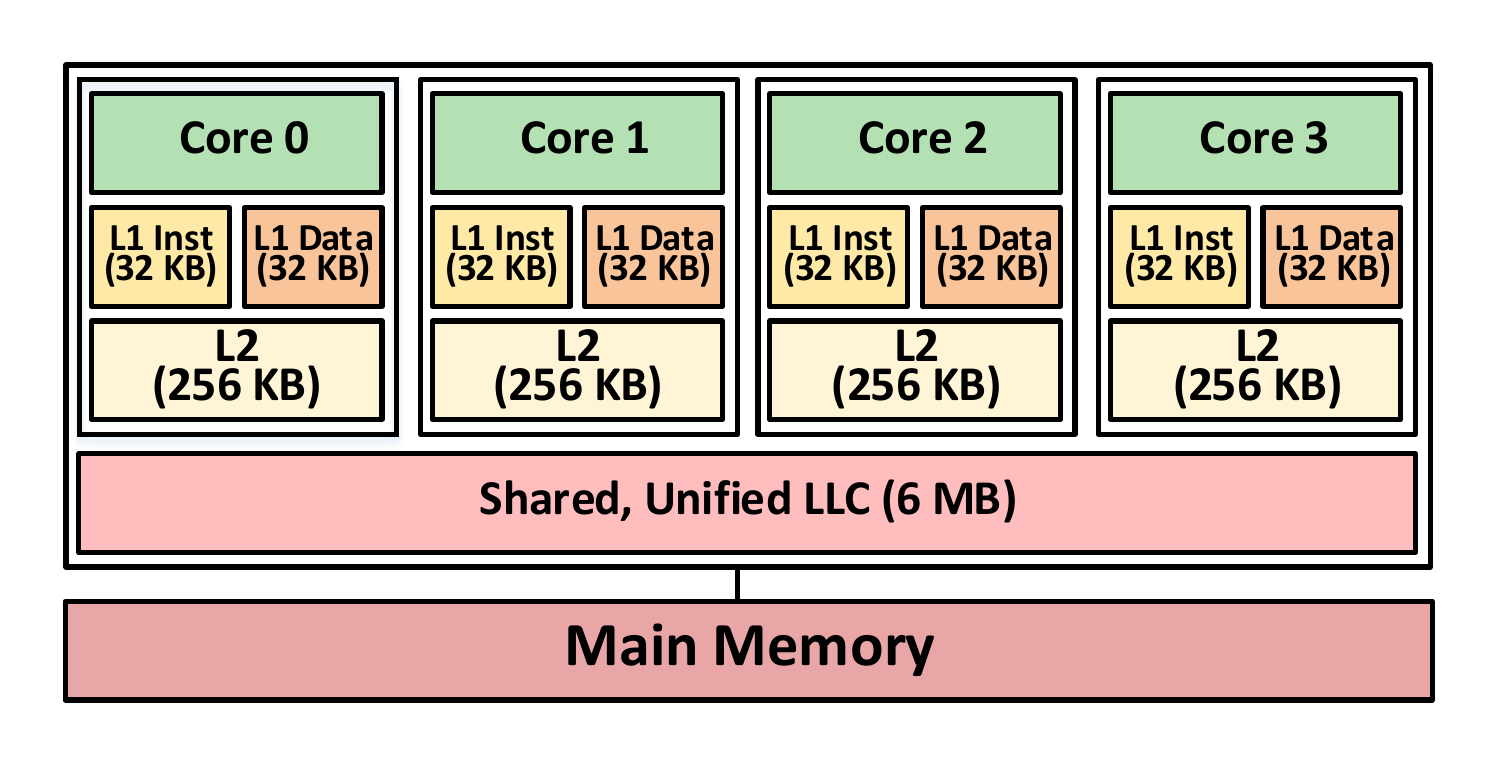}}
	\caption{The hierarchical structure of CPU cache.}
	\label{fig:cache}
	\end{figure}

	CPU cache exists at the top of the memory hierarchy.  A fast but small memory, employed to reduce the address translation overhead for most recently and frequently accessed data \cite{Cache:12}.  Data accessed by the processor is cached from the main memory with a \textit{cache line} granularity. In this context, cache hit and cache miss are defined as the existence and absence of a requested data in the cache, respectively. In today's commonly used processors, the cache consists of two per-core cache units, also known as $L1$ and $L2$, and the last level cache (LLC), which is shared among all cores. The shared characteristic of LLC has long been the cornerstone of the so-called cache attacks~\cite{Melt:4, Spec:5, FR:8}.
	Fig. \ref{fig:cache} shows an abstract representation of CPU cache.
	
	\subsection{Out-of-Order Execution}
	\label{subsec:out-of-order}
	
	In order to maximize the utilization of CPU's processing capacity, all available resources must be used exhaustively \cite{hennessy:13}. Such optimization requires different units in a CPU to work in parallel, in contrast with executing instructions sequentially. An instruction must be executed when all the required resources and dependencies are available, regardless of its position in the program \cite{hennessy:13}. The results obtained from this type of execution can then be rearranged in the original order and returned to the user.
	
\begin{figure}[t]
		\centerline{\includegraphics[width=1\columnwidth]{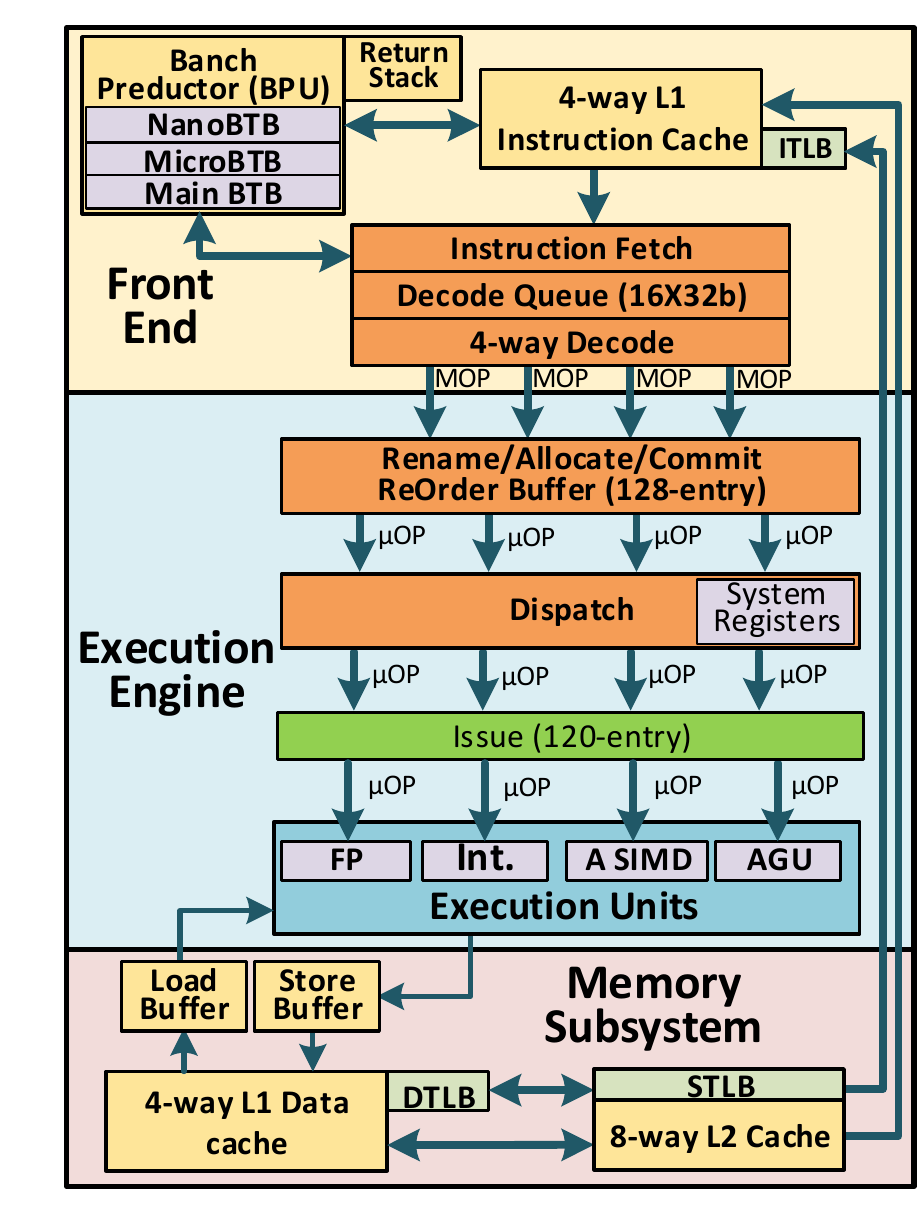}}
		\caption{The microarchitectural layout of a ARM cortex-A76 processor~\cite{a76}.}
		\label{fig:out}
	\end{figure}

	In many of modern CPUs, out-of-order execution is employed to significantly reduce the overhead associated with CPU units running idle, waiting for other units to complete their tasks. Thus, instructions far ahead in the instruction's sequence may be executed, if all the required dependencies are satisfied. In detail, as depicted in Fig. \ref{fig:out}, operations are broken down in to micro-operations ($\mu$-op), after being fetched in the frontend~\cite{a76}. The resulting $\mu$-ops are then executed in the execution engine. Instructions are then reassembled sequentially in reorder buffer, allowing them to be committed in their original order \cite{oor1:44}.
	
	\subsection{Branch Prediction}
	\label{subsec:branchprediction}
	
	The branch prediction mechanism is used to minimize the performance loss of the processor running idle while waiting for branch condition \cite{branch:14}. The branch prediction allows the processor to achieve its peak instruction throughput by providing an educated guess about a branch condition's outcome before the actual result is determined \cite{branch3:20}. 
	
	In order to provide a safe rollback point in case of a misprediction, the current state of the processor is saved. In case of a correct prediction, the speculatively executed instructions are committed, providing a huge performance advantage against running idle \cite{branch:14}. However, when a misprediction happens, the speculatively executed instructions are flushed from the pipeline~\cite{branchpr:42}. The branch prediction can also be used for indirect jumps \cite{branch3:20}, with the destination address not encoded in the instruction, and to be determined at the runtime. Records of previously taken branches, including the current program counter (PC) and previous branch destinations, are stored in a table called branch target buffer (BTB). The processor then uses BTB as a look-up table, to predict the return address of a branch destination~\cite{ branchpr:43}.
	
	\subsection{RSB}
	\label{subsec:RSB}
	
	In order to reduce idle time in processors, a function's return address must be available in a fast memory, before the actual return address is fetched from the software stack, located on the main memory. Therefore, modern processors employ RSB, a hardware located copy of the software stack, holding return addresses of the currently executing functions and providing them during the speculative execution \cite{ret:2}. In this manner, the processor can access the return address in RSB a few hundred cycles earlier but speculatively executed instructions are only committed if the predicted return address (RSB entry) matches the original return address (software stack entry).
	
	Due to its limited size, different microarchitectures have employed different mechanisms to handle overflows and underflows, by either switching to BTB or using RSB as a ring buffer and continue predicting \cite{ret:2}. To the best of our knowledge, AMD's processors stop predicting in case of an RSB overflow or underflow, while Intel's post-Skylake processors switch to branch predictor~\cite{ret:2}.
	The RSB size varies from 4 to 32 entries, in low-end machines to high-end servers, respectively.
	An underflow can happen when multiple nested calls push their return addresses into RSB, resulting in older entries to be overwritten due to the size limit. Therefore, as the nested calls return and their associated RSB entries are popped out, original entries for outer functions are no longer available, as RSB is empty \cite{ret:2, RSB:1}. 
	 Like other optimization and performance-enhancing mechanisms, RSB has been the source of many attacks. An attacker can trigger many scenarios, in which RSB can be exploited to misdirect the victim's control flow. These strategies can be utilized to subvert isolation principals in CPU and operating system \cite{ret:2, RSB:1, ROP:59}. In the following, we explain two strategies used to perform attacks on RSB: 
	 \begin{itemize}
	 	\item \textbf{Direct Miss predict:}
	 	Executing call instructions can arbitrarily push entries in RSB, resulting in the overwriting victim's entries. Therefore, a misprediction is triggered when the processor speculatively returns to the overwritten value \cite{ret:2, RSB:1}.
	
	 	\item \textbf{Speculative Miss predict:}
	 	Speculatively called functions also push their return addresses into RSovB. Although the speculatively executed instructions are unrolled in case of a misprediction, speculative entries in the RSB are not discarded \cite{ret:2, RSB:1}.
	 \end{itemize}
	
	\subsection{Covert Channels}
	\label{subsec:covert channels}
	Covert channels allow two or more parties to secretly communicate with each other, bypassing isolation boundaries enforced by the operating system or CPU. Similar to side-channel attacks, covert channels pose various threats to computers' security. Studying the possibility of potential covert channels in various layers of a system has been a long quest for security researchers \cite{cov1:64, cov2:65}.
	
	Previous work exploits various shared resources, such as cache and BTB to implement covert channels. Gruss \textit{et al.,} shows that the Flush+Reload \cite{FR:8} and the Flush+Flush~\cite{FF:67} attack can be used to implement a cache-based covert channel. These attacks introduce effective covert channels based on commonly targeted system components.
	Recently, Chakraborty \textit{et al.} \cite{tches-2021-30800} proposes a covert channel based on RSB. In this method, the receiver first fills the RSB, then the sender executes a function or does nothing to push or not to push an irrelevant address to the RSB structure to transfer 1 or 0 to the receiver. Next, in the receiver process, CPU fetches the address from RSB to speculate the return address. Based on the run time of these functions, the receiver determines the transmitted bit. When the sender injects an address to the filled RSB, the oldest address is removed from RSB and this increases the run time of the receiver process. This covert channel can achieve up to 30 KB/sec bandwidth with 75\% to 85\% accuracy.

	\section{Implemented Attacks Overview}
	\label{sec:attacks}
	
	To study and analyze ARM processors security, we review and discuss corresponding attacks that have been implemented for this work.
	Transient execution attacks have different variants~\cite{m3:36}. In this work, we focus on the most common variants of these attacks that have been used in related works. Therefore, we first explain \mbox{Spectre-PHT (Variant 1)}, which is the base Spectre attack, then we explain Meltdown (Variant 3), Meltdown-GP (Variant 3a), Spectre-STL (Variant 4), and SpectreRSB. The Meltdown-GP (Variant 3a) and SpectreRSB attacks, are attacks based on RSB and have not been implemented on ARM processors earlier.
	We note that due to the similar procedure of SpectreRSB and Spectre-BTB (Variant 2)~\cite{arm:cachespeculation}, in this manner that SpectreRSB is assumed as an extension for the Spectre-BTB, we have only implemented SpectreRSB in this work. The evaluation of Spectre-BTB and other transient execution attacks on ARM can be done as part of the future work.
	
	
	
	
	
	
	\subsection{Spectre-PHT (Variant 1)}
	\label{sebsec:ver1}
	
	
	In this variant, the attacker miss-trains the branch predictor to decide a branch instruction with a designated result, causing the speculative execution of code regions protected by a bound check \textit{if} as well as other security precautions. 
	
	The attacker achieves this goal by executing the targeted branch instruction with a legal condition value such that the branch condition becomes \textit{True}. The branch predictor is therefore trained to assume a \textit{True} result for that specific branch. This results in assigning \textit{True} to that branch instruction during the speculative execution, even though the branch input is maliciously manipulated by the attacker to access illegal or out-of-bound memory \cite{Spec:5}. The speculatively accessed memory would trigger a cache hit in the retrieval phase, leaking the victim’s secret consequently \cite{FR:8, llcA:7}.

	
	\subsection{Meltdown (Variant 3/3a)}
	\label{sebsec:ver3}
	
	In Variant 3 (aka Meltdown) and Variant 3a, the attacker exploits speculative execution to disclose the data. In normal situation, attacker's attempts to access this data are rejected by an exception from the CPU. In variant 3 and variant 3a, the attacker accesses the kernel memory region and the CPU registers, respectively, using an unprivileged process such as one running in the user mode. Note that some critical information regarding CPU operation is stored on these specific registers, which are not even accessible in privileged mode. \cite{arm:tja1043}.
	On a mistaken speculative execution, speculatively-executed instruction are reverted and flushed form the CPU pipline and the program continues in the valid direction. However, speculative execution can leak the victim's information to the CPU cache. In this paper, we perform the Variant 3a attack using the SpectreRSB technique.
	
	
	\subsection{Spectre-STL (Variant 4)}
	\label{sebsec:ver4}
\begin{figure}[htbp]
		\centerline{\includegraphics[width=1\columnwidth]{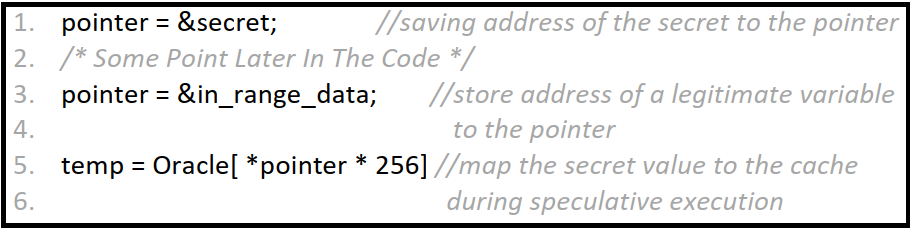}}
		\caption{Simplified variant-4 attack scenario.}
		\label{ssb}
	\end{figure}

	The fourth variant of this vulnerability is known as Store To Load or STL. In this variant, the attacker exploits the memory disambiguation predictors feature. The memory disambiguation predictors are used in high-performance CPUs to allow load instructions to be executed speculatively even when the previous store instruction address is unknown. In the case of overlapping the preceding store instruction's address with the load command, the CPU will re-execute the load command with the updated address. Fig.\ref{ssb} shows the simplified code for this attack. In line 1, the attacker stores the address of the \textit{secret} in the \textit{pointer}. In line 3, the address of a legitimate variable is stored in the pointer. Then, the attacker loads the pointer address (line 5).
	Because of the memory disambiguation feature in the CPUs, the CPU may execute line 5 before evaluating the new value of the pointer. In this situation, line 5 will be executed out-of-orderly while the address of the secret is stored in the pointer. Consequently, the secret will be mapped to the cache, which can be exploited with the Flush+Reload attack.
	When CPU computed the new value of the \textit{pointer}, it will discard the result of the out-of-order executed instructions and re-execute line 5 with the new address.

	\subsection{SpectreRSB}
	\label{subsec:spectreRSB}
	
	Instead of exploiting the branch prediction mechanism, to perform illegal access to the victim’s sensitive information, SpectreRSB \cite{RSB:1} exploits RSB.
	 When the speculative execution reaches a return instruction, instead of waiting for the return address to be fetched from the main memory~\cite{SpecExe:15}, RSB is looked up to obtain the return address and continue the speculative execution. Finally, when the original return address is fetched from the software stack, the processor then matches the RSB return address to the original one. It accordingly decides whether to commit or to squash the speculatively executed instructions, in case of a match or mismatch respectively~\cite{branch3:20}.
	 With RSB being out of the attacker's reach, the attacker's goal is to manipulate the software stack such that it would trigger a misprediction, preventing the processor from committing the illegally executed instruction and causing an exception \cite{RSB:1, ret:2}.
	
\begin{figure}[htbp]
		\centerline{\includegraphics[width=1\columnwidth]{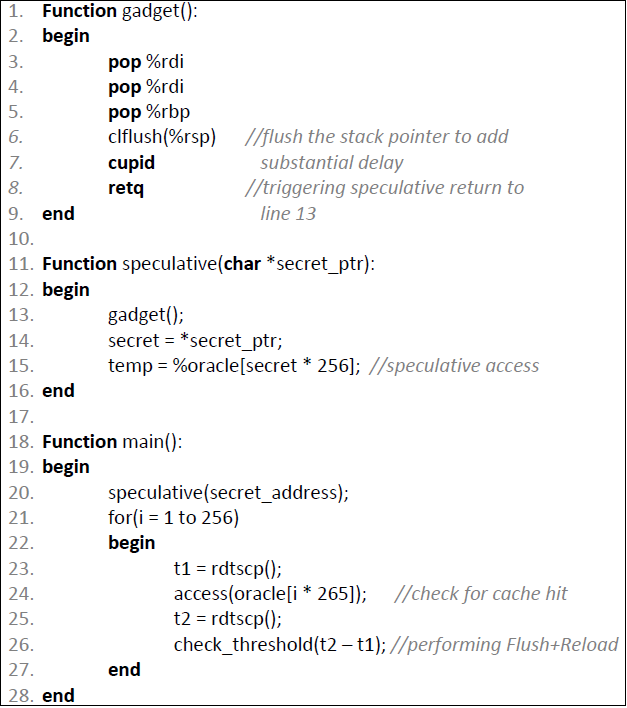}}
		\caption{Simplified SpectreRSB attack scenario \cite{RSB:1}.}
		\label{rsb}
	\end{figure}

	Fig. \ref{rsb} shows the pseudo code for SpectreRSB attacks.
	As shown in the figure, the attacker calls the gadget function, which manipulates the software stack such that it would become inconsistent with RSB (lines 1-9). The processor, then uses RSB to speculatively return to an illegal memory address. The resulting speculative execution makes the victim's secret available on the cache (lines 14 and 15). Finally, when the manipulated return address is fetched, which results in a mismatch with the original RSB return address, the processor squashes the speculatively executed instructions and continues the program from line 21. However, the cached data is still present, and then the attacker can retrieve victim's secret from cache with timing side channel attacks such as Flush+Reload (line 21-27). Although this attack has been implemented on Intel processors, as mentioned in related work, it has not been investigated on ARM processors \cite{RSB:1, ret:2}. 
	
	\subsection{Flush+Reload Attack}
	\label{subsec:flush+reload}

	
	The Flush+Reload attack consists of three main phases. In the initial phase, the attacker flushes the monitored cache lines. Secondly, the attacker waits for the victim to perform a memory read operation. In the last phase, the attacker accesses the monitored cache line. In this step, the attacker can determine if that specific memory line was accessed by the victim, by measuring the time it takes for the monitored memory line to be fetched \cite{FR:8}. If the victim had accessed the monitored memory line, a cache hit is triggered, and consequently, the access time for the attacker would reduce significantly. In contrast, if the victim had not accessed the monitored memory line, the resulting cache miss, would signal the attacker about the victim's behavior.
	The Flush+Reload attack has been widely used in microarchitecture attacks to read speculative data from cache \cite{Melt:4, Spec:5, RSB:1, ret:2}.

	\section{Attacks Building Blocks}
	\label{sec:building blocks}
	
	In contrast with Intel processors and previous straightforward approaches towards implementing cache attacks, microarchitectural diversity amongst various series of ARM processors, poses many complexities and difficulties regarding implementing these attacks. 
	In this section, we present and discuss major recurring building blocks, i.e., timer and eviction strategy, that play a crucial role in the implementation of side-channel attacks on ARM processors. To implement attacks in different scenarios, we introduce these building blocks in high and low privileges.
	\begin{table}[t]
		\centering
		\caption{Best eviction variables for different ARM Cores}
		\label{tab:eviction}
		\begin{threeparttable}
			\begin{tabular}{|l|c|c|c|}
				\hline
				\multicolumn{1}{|c|}{\multirow{2}{*}{\textbf{CPU Core}}} & \multicolumn{3}{c|}{\textbf{Eviction Strategy}} \\ \cline{2-4} 
				\multicolumn{1}{|c|}{}                          & \textbf{N}\tnote{$\dagger$}           & \textbf{A}\tnote{$\bullet$}          & \textbf{D}\tnote{$\diamond$}          \\ \hline
				Cortex-A53                                      & 21          & 2          & 5           \\ \hline
				Cortex-A8                                       & -           & -          & -           \\ \hline
				Cortex-A9                                       & 10          & 3          & 6           \\ \hline
				Cortex-A72                                      & 7           & 1          & 16          \\ \hline
			\end{tabular}
			\begin{tablenotes}
				\item[$\dagger$] Determines the length of loop
				\item[$\bullet$] Determines shift offset
				\item[$\diamond$]Number of memory access in each iteration
			\end{tablenotes}
		\end{threeparttable}
	\end{table} 
	
	\begin{table*}[t]
		\centering
		\footnotesize
		\caption{Specification Of Processors Used In Our Experiments }
		\label{tab:test setup}
		\begin{tabular}{|l|c|c|c|c|c|c|c|}
			\hline
			\multicolumn{1}{|l|}{\textbf{CPU Core}} & \textbf{SOC}     & \textbf{Instruction set architecture} & \textbf{Pipeline} & \textbf{$L1$ I-Cache / D-Cache} & \textbf{$L2$ Cache} & \multicolumn{1}{l|}{\textbf{L3 Cache}} \\ \hline \hline
			Cortex-A53                              & Broadcom BCM2837 & ARMv8-A                               & In order          & 32KB / 32KB                   & 512KB             & NA                                     \\ \hline
			Cortex-A8                               & TI AM335x        & ARMv7-A                               & In order          & 32KB / 32KB                   & 256KB             & NA                                     \\ \hline
			Cortex-A9                               & Zync-7000        & ARMv7-A                               & Out of order      & 32KB / 32KB                   & L2C-310 512KB     & NA                                     \\ \hline
			Cortex-A72                              & Broadcom BCM2711 & ARMv8-A                               & Out of order      & 48KB / 32KB                   & 1MB               & NA                                     \\ \hline
			Core i7-4500u                           & -                & X86-64                                & Out of order      & 32KB / 32KB                   & 256KB             & 4096KB                                 \\ \hline
			
		\end{tabular}
	\end{table*}

	\subsection{Timer}
	\label{subsec:PMU}
	
	A high-resolution timer is an essential building block, to implement microarchitectural attacks. On Intel's processors, the time sampling in the Flush+Reload attack is performed through the \textit{rdtsc} instruction, which returns the current CPU cycle count. However, on ARM processors, a different approach is required to measure time. In the following, we explain some of the methods used in previous work to implement timers in high or low privileges.

	\textbf{Performance monitoring unit (PMU):} This unit is nowadays found in most modern processors. As an on-chip hardware unit, the PMU allows the user to monitor the microarchitectural state of the CPU. Providing a wide range of information such as instruction cycle count, cache hit, cache miss, and branch prediction results, the PMU helps analyze CPU's behavior when a program is being executed. 
	the PMU consists of model-specific registers (MSR), which can be configured to store different performance parameters. We should mention that direct access to the PMU is not available in the userspace processes by default. Therefore, to obtain our implementations, the PMU has been enabled temporarily in our test system by executing privileged instruction.
	
	\textbf{Dedicated timer:} The attacker can create a counter thread, which increases a counter in an infinite loop, and uses the counter's value to measure time \cite{ARMag:3}. This timer runs in userspace, and the resolution of this timer is enough to distinguish between a cache hit and cache miss.
	


	\begin{table*}[t]
		\centering
		\footnotesize
		\caption{SpectreRSB Implementation Result}
		\label{tab:spectreRSB}
		\begin{tabular}{c|c|c|c|c|c}
			\cmidrule[1pt](lr){1-6}
			\multirow{2}{*}{CPU / SOC} & \multirow{2}{*}{Speculative Load} & \multicolumn{2}{c|}{SpectreRSB when secret resides in $L1$} & \multicolumn{2}{c}{SpectreRSB when secret resides in main memory} \\ \cmidrule(lr){3-6} 
			&  & Cache Miss & Page Fault & Cache Miss & Page Fault \\ \cmidrule[1pt](lr){1-6}
			Cortex-A53 / BCM2837 & \ding{55} & \ding{55} & - & \ding{55} & - \\ 
			Cortex-A8 / AM335x & \ding{55} &   \ding{55} & - & \ding{55} & - \\ 
			Cortex-A9 / ZYNQ7000 & \ding{51} & \ding{55} & - & \ding{55} & - \\ 
			Cortex-A72 / BCM2711 & \ding{51} & \ding{51} & - & \ding{55} & - \\ 
			Core i7-4500u / - & \ding{51} &    \ding{51} & - & \ding{51} & - \\ 
			\cmidrule[1pt](lr){1-6}
		\end{tabular}
	\end{table*}

	\subsection{Eviction Strategy}
	\label{subsec:eviction}
	
	In order to maximize the required speculative execution window for the leaked data to be mapped into the cache, the access time for the original return address must be increased. This avoids the speculatively executed instructions to be determined as a mistakenly taken branch, immaturely \cite{FR:8, Spec:5, Melt:4}. To achieve this, we must evict the stack pointer from the cache \cite{ret:2, c1:22}. Not limited to the stack pointer, in different experiments and attack scenarios, it is required to evict different data from the cache. These experiments are introduced and explained in detail in Section \ref{sec:evaluation}.
	
	Unlike Intel processors, which provide the user with the unprivileged \texttt{clflush} instruction to flush arbitrary data from the cache \cite{Melt:4}, no such unprivileged mechanism is incorporated in ARM processors~\cite{ARMag:3}. In this work, after obtaining proof of concept for cache eviction on ARM processors through native instructions, we implemented the solution proposed by Lipp \textit{et al.}, \cite{ARMag:3} to evict data from the cache in unprivileged mode. As each processor model in the ARM family incorporates a different microarchitectural design, we test different eviction strategy on each ARM processor. \mbox{Table \ref{tab:eviction}} shows eviction variables, calculated in our work for each processor, based on the approach introduced in~\cite{ARMag:3}. These variables determine the loop parameters to successfully evict data from the cache.
	We should point out that we could not find an effective eviction strategy for ARM Cortex-A8 CPU.
	This is because Cortex-A8 appears to follow different cache address-bit mapping from the standard one. 
	Therefore, to effectively evict the target address from the Cortex-A8 cache in the user mode, we access a large structure that is 3$\times$ of cache size~\cite{spreitzer2013cache}. However, this approach has some downside compared to the ARMagedon approach, which is requiring more time to complete. Moreover, to obtain our proof of concept implementations, we have used privileged cache eviction instructions, provided by ARM.
	
	\subsection{Triggering a Page Fault}
	\label{subsec:PageFault}
	
	A page fault would be triggered during access to a memory location that has not been mapped to the virtual address space of a process by the memory management unit (MMU). We note that resolving a condition that triggers a page fault would take longer than resolving a condition that triggers a cache miss. Consequently, exploiting the substantial delay can effectively extend the speculative window. The required page fault can occur during access to an uninitialized variable, or by manually modifying a process's page table. Manually modifying the page table requires root privilege and should be done in kernel space. In our experiments, we trigger a page fault by accessing an uninitialized variable in user mode. 
	
	\section{Evaluation}
	\label{sec:evaluation}
	We implemented transient execution attacks (Spectre, Meltdown, and SpectreRSB) introduced in Section~\ref{sec:attacks} on various ARM and Intel CPUs.  
	Our experiments targeted a wide range of ARM processors, from a low-end processor such as Cortex-A8 to a high-end processor like Cortex-A72 on Raspberry Pi 4, which has been released after the identification of Spectre-family vulnerabilities. We selected these processors to evaluate the susceptibility of various ARM microarchitectures (32/64 bit instruction size, in-order/out-of-order execution, ARM-V7/ARM-V8) released at different times.
	Moreover, to achieve a better evaluation, we performed our experiments on Intel Core i7-4500u. Model and attributes of CPUs investigated in our experiments are shown in Table~\ref{tab:test setup}.
	
	In the following sections, we first explain and clarify our experiments' conditions and then we provide our experimental results for the SpectreRSB, Variant 1, 3, 3a, and 4 attacks on the ARM processor. Finally, we discuss on experimental result and conclude the result that is obtained from our experiments.
			\begin{table*}[t]
		\centering
		\caption{Spectre (Variant 1) Implementation Result}
		\label{tab:spectre}
		\footnotesize
		\begin{tabular}{c|c|c|c|c|c}
			\cmidrule[1pt](lr){1-6}
			\multirow{2}{*}{CPU / SOC} & \multirow{2}{*}{Speculative Load} & \multicolumn{2}{c|}{Spectre when secret resides in $L1$} & \multicolumn{2}{c}{Spectre when secret resides in main memory} \\ \cmidrule(lr){3-6} 
			&  & Cache Miss & Page Fault & Cache Miss & Page Fault \\ \cmidrule[1pt](lr){1-6}
			Cortex-A53 / BCM2837 & \ding{55} & \ding{55} & \ding{55} & \ding{55} & \ding{55} \\ 
			Cortex-A8 / AM335x & \ding{55} &   \ding{55} & \ding{55} & \ding{55} & \ding{55} \\ 
			Cortex-A9 / ZYNQ7000 & \ding{51} & \ding{55} & \ding{51} & \ding{55} & \ding{51} \\ 
			Cortex-A72 / BCM2711 & \ding{51} & \ding{51} & \ding{51} & \ding{51} & \ding{51} \\ 
			Core i7-4500u / - & \ding{51} &    \ding{51} & \ding{51} & \ding{51} & \ding{51} \\ 
			\cmidrule[1pt](lr){1-6}
		\end{tabular}
	\end{table*}
	\subsection{Attack Scenario}
	\label{subsec:scenarios}
	
	Before discussing our experimental results, we should clarify these experimental scenarios and attack models. To achieve a deeper and more precise evaluation, all of our attacks are implemented and analyzed in the following scenarios. In these experiments, in addition to implementing these attacks, we aim to evaluate the sensitivity of each ARM core against these attacks.
	To explore the possibility of mistraining the CPU branch predictor unit to run arbitrary code during the speculative execution window, we have designed and implemented \textit{Speculative Load} scenario (column 2 in table~\ref{tab:spectre}, \ref{tab:spectreRSB}). We achieved this by executing the smallest possible gadget during the speculative window, \textit{e.g.}, loading uncached data into the cache, which can be done with a single assembly instruction. Whereas, in order to implement the spectre-family attacks, the attacker needs to execute four assembly instructions ((1) load the secret from memory to the CPU's register; (2) multiply it with cache line size, e.g., 64; (3) add the result to the oracle base address; (4) load the resulted memory address to the cache) to cache the secret data during the speculative execution. Also, this window can be extended so that the attacker has more time to access the secret data. The processor is more likely to be susceptible to these attacks if the attacker retrieves the secret data with a smaller window size.
	In our experiments, to extend the speculative window, we have designed two below scenarios.
	\begin{itemize}[leftmargin=0.1in]
		
		\item\textbf{Cache Miss:} 
		In this scenario, to extend the speculative execution window, we have implemented the attacks once a cache miss occurs during an access to the \emph{if} operand or the function return address. Because of the cache miss occurrence, CPU needs a longer time to evaluate the result of the conditional operand or the return addresses. Therefore, the attacker has more time to access the secret speculatively.
		
		\item\textbf{Page Fault:} In this scenario, we further extend the speculative window compare to the cache miss scenario. To this end, we can use a memory location that is accessible to the attacker process but currently is not mapped by the MMU into the attacker's virtual address space (see Section~\ref{subsec:PageFault}). Therefore, resolving the condition would take longer, and consequently, the gadget would have more chances to execute entirely in comparison to the cache miss scenario.
	\end{itemize}
	Moreover, the victim's secret, which the attacker wants to access, can reside in either the $L1$ cache or the main-memory (DRAM). A larger speculative window is required when the secret resides in the main memory. Therefore, if the attack is implemented in this scenario, the processor is more sensitive to the Spectre vulnerabilities.
	Note that we implemented all attacks with these scenarios on ARM processor with C and Inline assembly.
	
	

	\subsection{SpectreRSB}
	\label{subsec:spectrersb}
	
	To further explore vulnerabilities and design flaws in the indicated CPUs, we present an implementation of SpectreRSB attack and discuss its effectiveness in both speculative load and cache miss scenarios. Our results are shown in Table \ref{tab:spectreRSB}. It is necessary to point out that due to the frequently accessed nature of the stack, 
	we couldn't remove the return address from the memory and trigger a page fault. Therefore, all the SpectreRSB attack results are based on the cache miss scenario.


	Our results demonstrate that ARM Cortex-A72 is vulnerable to SpectreRSB attack, when the secret resides in the $L1$ cache. However, our attempts did not yield any results when the secret resides in the main memory.  
	Similarly, due to the small speculative window size, we could not implement this attack on ARM Cortex-A9 in any of the scenarios. Furthermore, due to the in-order pipeline used in Cortex-A8 and Cortex-A53, these processors are not susceptible to this attack. 
	Our results demonstrate that SpectreRSB attacks can pose a severe threat to ARM processors and make a significant impact on devices incorporating these CPU models. 
	Proof-of-concepts obtained in our experimental analysis can be used to create realistic and remote attack scenarios as well. Moreover, we managed to successfully implement this attack on Intel processor, in both scenarios. 
	
	\subsection{Spectre (Variant 1)}
	\label{subsec:spectre}
	
	Table~\ref{tab:spectre} shows the implementation results of the \mbox{Spectre V-1} (aka bound check bypass) attack for the described scenarios. As shown in Table \ref{tab:spectre}, in Cortex-A8 and Cortex-A53, our attempts did not yield any meaningful results. This is due to incorporating an in-order pipeline in these processors which is determined by speculative load scenario (column 2 in table~\ref{tab:spectre}).
	In Cortex-A9 we could successfully test the  Spectre attack in page fault scenarios when secret resides in the $L1$ and memory; however, we were unable to implement the cache miss scenario.
	Moreover, as shown in Table~\ref{tab:spectre}, all scenarios have been successfully implemented on Cortex-A72 on Raspberry Pi 4. As expected, we successfully implemented this attack in all scenarios, on the Intel processor without any difficulty.
		\begin{table}[t]
		\centering
		\caption{Variant 3, 3a, and 4 Implementation Result}
		\label{tab:evaluation}
		
		\begin{tabular}{l c c c}
			\cmidrule[1pt](lr){1-4}
			\diagbox[width=12em]{\textbf{CPU}}{\textbf{Attack}}
			& \textbf{V 3} & \textbf{V 3a} & \textbf{V 4} \\ \cmidrule(lr){1-4}
			\textbf{Cortex-A53 / BCM2837} & \ding{55} & \ding{55} & \ding{55} \\
			\textbf{Cortex-A8 / AM335x} & \ding{55} & \ding{55} & \ding{55} \\ 
			\textbf{Cortex-A9 / ZYNQ7000} & \ding{55} & \ding{55} & \ding{55} \\ 
			\textbf{Cortex-A72 / BCM2711} & \ding{55} & \ding{51} & \ding{51} \\
			\textbf{Core i7-4500u / -} & \ding{51} & \ding{51} & \ding{51} \\ 
			\cmidrule[1pt](lr){1-4}
		\end{tabular}
	\end{table}
	
	\subsection{Meltdown (Variant 3/3a)}
	\label{subsec:spectre_3a}
	In this experiment, we describe the implementation result of the Variant 3/3a on the CPUs mentioned in Table~\ref{tab:test setup}. As described in Section~\ref{sec:attacks}, Meltdown enables the attacker to access data that are not expected to be accessible at the current exception level. The target data in Variant 3 is the kernel memory, while it is the special CPU registers in Variant 3a. Our experiments confirm the resistance of ARM processors against the Meltdown attacks~\cite{Melt:4}. However, we have successfully exploited Variant 3a to access the CPU registers in Cortex-A72, which is used in Raspberry Pi 4. The result for this attack is shown in Table \ref{tab:evaluation}.
	Also, we should mention that we implemented this attack with SpectreRSB vulnerabilities. Herein, instead of \textit{load} instruction which loads a memory slot to a CPU's register, we've used \textit{MRC} instruction which loads a special register to a CPU's register.
	
		\begin{figure*}
		\includegraphics[width=1.05\textwidth]{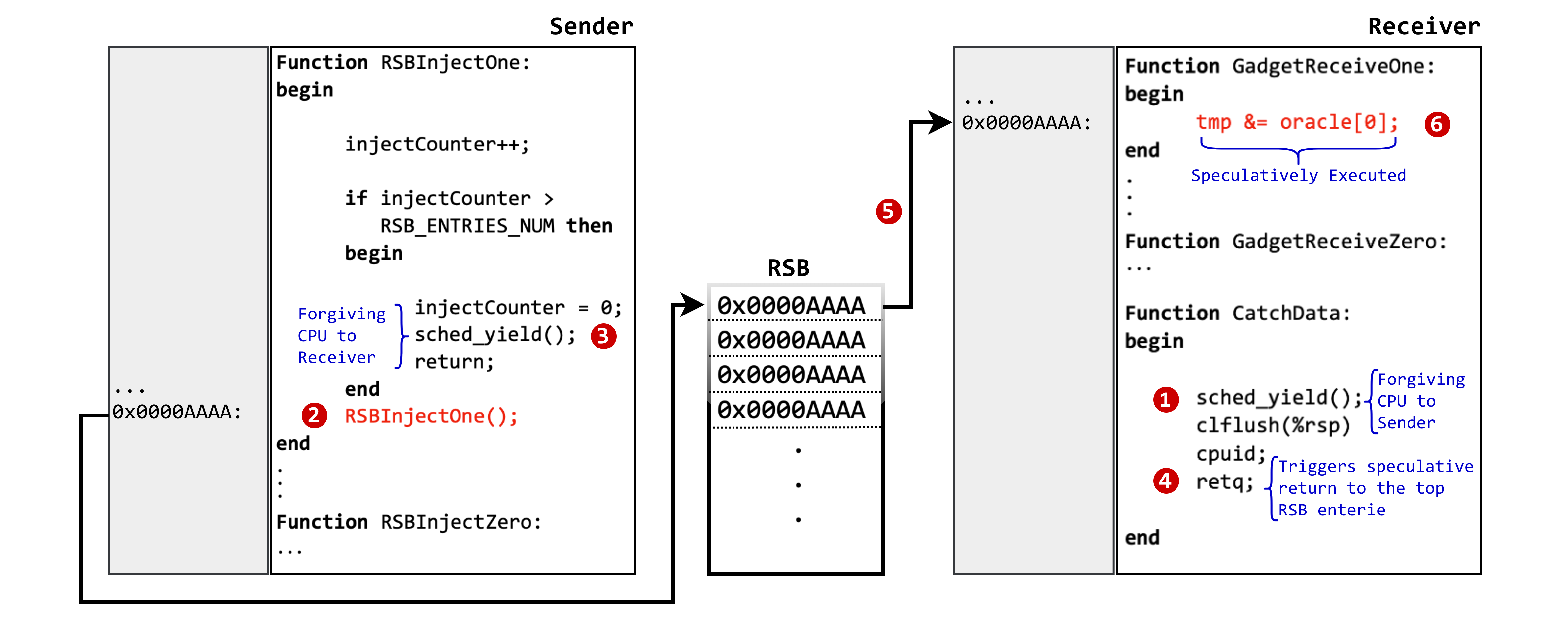}
		\centering
		\caption{Exploiting RSB as a shared resource to initiate a covert channel.}
		\label{pic:covertchannel}
	\end{figure*}
	\subsection{Spectre (Variant 4)}
	\label{subsec:spectrev4}
	
	As described in Section \ref{sebsec:ver4}, variant 4 exploits the fact that the modern CPUs may speculate a load address and consequently load the speculated memory address to the cache in order to access memory locations that should not be accessed. Table~\ref{tab:evaluation} shows the implementation results of this variant. As shown in Table~\ref{tab:evaluation}, Cortex-A8/A9/A53 are not vulnerable to this version, , while Cortex-A72 is susceptible when the secret data resides in the L1 cache.
	Note that, successfully mounting this attack requires an assembly program.

	\subsection{Discussion}
	\label{subsec:discussion}
	In this section, we investigate the transient execution attacks vulnerability on different ARM processors. In order to achieve a more precise evaluation, we implement these attacks in different scenarios. The results show that ARM Cortex-A8 and Cortex-A53 are not vulnerable to Spectre attacks due to their \textit{in-order} pipeline. In these processors, it is not feasible to access uncached data during the speculative window (\textit{Speculative Load} scenario). However, Cortex-A72 and Cortex-A9 are susceptible to Spectre attack V-1, and also, we have successfully implemented SpectreRSB on Cortex-A72. Moreover, although Cortex-A9 supports speculative execution and is vulnerable to the Spectre attack, since we could not implement the cache miss scenario on Cortex-A9, we can deduce that this processor has a shorter speculative window. Therefore, Cortex-A9 is less vulnerable to these attacks compared to Cortex-A72.
		We have implemented the Spectre V-1, V-3a, and V-4 on Cortex-A72, which is used in Raspberry Pi 4. Since  Raspberry Pi 4 has been released after disclosing this security flaw, our results indicate that the CPU manufacturers have neglected this vulnerability.
	Our experiments show that, unlike SpectreRSB, we can implement Spectre V-1 on Cortex-A72 when the secret data resides in the main memory. So we can conclude that the Spectre V-1 needs less speculative window size and is a more serious threat for the processors.
		Also, our experiments on Meltdown vulnerability confirm the previous work~\cite{Melt:4} that the ARM processors are not vulnerable to this attack. Nevertheless, we have successfully implemented Meltdown variant 3a on Cortex-A72.
	
	\section{Covert Channel}
	\label{sec:covert channel}
	
	In this section, we introduce a new noise-free covert channel, based on the SpectreRSB attack.
	By exploiting design flaws in RSB, first we demonstrate how two malicious processes can establish a noise-free covert channel and subvert cross-process boundaries. Secondly, we further provide an evaluation of the proposed covert channel.
	
	As we investigated the SpectreRSB attack on various processors, we noticed that RSB, shared between processes on the same physical core, offers an effective environment for unauthorized communication between malicious processes. Moreover, we introduce a method to increase bandwidth and reduce the noise of the covert channel. To implement our proposed covert channel, we have the following assumptions:
	
	\textbf{Physical core co-residence:} 
	Since current processors incorporate a separate RSB for each physical core, the sender and the receiver processes should reside on the same physical core to successfully establish a covert communication channel. The same pre-requirement also has been considered in the previous work \cite{BRCOV:68, BRCOVUP:69}, where they have introduced covert channels through the gShare branch predictor; so the covert channel is established in one physical core.
	
	\textbf{Post-Spectre security patches:}
	After finding the Spectre attacks, a security patch has been released for the operating systems. Hence, in this work, similar to \cite{RSB:1, ret:2}, we disable Post-Spectre security patches temporarily.
	
	\textbf{ASLR:} 
	The default settings of ASLR (address space layout randomization) on operating systems would not pose any difficulty for our covert channel.

	\textbf{Resides in the same core:}
	Pinning two processes on the same core can be easily achieved with the normal APIs and does not need any special privilege.

	\subsection{Implementation Details}  
	\label{subsec:implementation details}
	
	In our approach towards establishing a covert channel, unlike previous work, we used two separate address locations (say $x$ and $y$) instead of one location. The cache hit for $x$, for example, indicates a \textit{zero} value and the cache hit for $y$ indicates a \textit{one} value. With this approach, we can significantly reduce the noise, and by further increasing the address locations, we can increase bandwidth to $2\times$ or more. 
	In the following, we explain the procedure of the sender and receiver in our covert channel to transfer a single bit. These steps are depicted in Fig.~\ref{pic:covertchannel}.
	
	\begin{itemize}[leftmargin=0.1in]
		\item \textbf{Sender:}
		First, the sender injects the return address for one of the two gadgets into RSB by repeatedly calling the gadget within itself. Gadgets \texttt{RSBInjectOne} and \texttt{RSBInjectZero} represent a \textit{one} or a \textit{zero} value, respectively. Note that by increasing the number of gadgets and memory locations, we can transfer more bits per each context switch, and consequently, increase the bandwidth effectively. Next, when the latest RSB entries are filled with the gadget's address, the sender yields the control to the receiver (executes \texttt{sched\_yield()}). 
		
		\item \textbf{Receiver:}
		After a context switch, the receiver executes the next instructions until it reaches the \texttt{retq} instruction. Next, the process speculatively executes the latest return address in RSB. The secret data is speculatively accessed in the gadget, causing an element in the oracle to be cached. Now, the receiver performs the Flush+Reload technique on the oracle, measuring the time it takes for each oracle's element to be fetched. The receiver catches a \textit{one} value if the \texttt{oracle[0]} is present in the cache, and similarly a \textit{zero} value if the \texttt{oracle[64]} is present in the cache. Finally, the receiver yields the control to the sender to receive the next bit.

	\end{itemize}

	It should be noted that both gadgets should reside on the exact same virtual addresses in the sender's and receiver's address spaces. However, this can be easily achieved by \textit{Just-In-Time Compiling (Jitting)} or applying linker configurations at the compile time.
	
	\subsection{Performance Evaluation}
	\label{performance evaluation}
	
	\begin{table}[]
		\centering
		\caption{Covert channel's speed and required memory}
		\label{tab:covertchannel}
		\begin{tabular}{|c|c|c|}
			\hline
			\textbf{\begin{tabular}[c]{@{}c@{}}Bit Per\\  Context Switch\end{tabular}} & \textbf{\begin{tabular}[c]{@{}c@{}}Speed \\ (KB/sec)\end{tabular}} & \textbf{\begin{tabular}[c]{@{}c@{}}Required Memory \\ (Byte)\end{tabular}} \\ \hline
			1                                                                          & 52.92                                                              & 128                                                                        \\ \hline
			2                                                                          & 84.63                                                              & 256                                                                        \\ \hline
			\textbf{3}                                                                 & \textbf{94.19}                                                     & \textbf{512}                                                               \\ \hline
			4                                                                          & 88.88                                                              & 1024                                                                       \\ \hline
			5                                                                          & 66.2                                                               & 2048                                                                       \\ \hline
			6                                                                          & 43.35                                                              & 4096                                                                       \\ \hline
		\end{tabular}
	\end{table}
	We test our covert channel on a PC with core i7-4500u CPU using Fedora workstation with kernel version 5.5.10.
	\begin{figure*}[htb]
		\includegraphics[width=\textwidth]{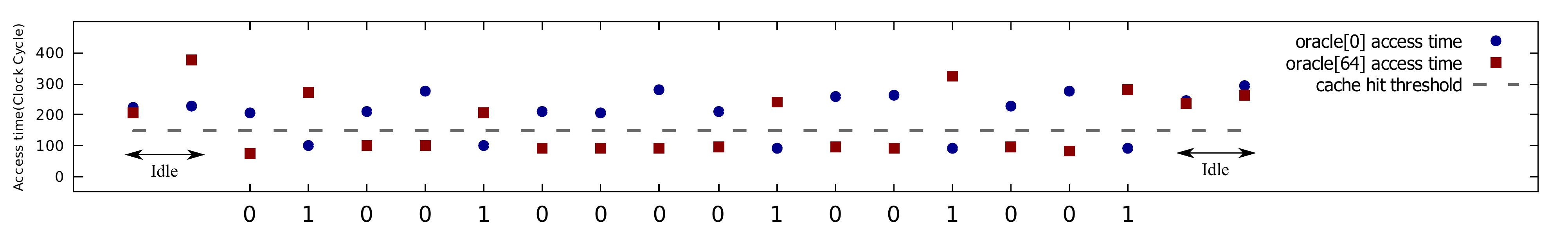}
		\centering	
		
		\caption{Measured access times for the \texttt{Oracle} during transferring \textit{"HI"} with ascii code \textit{0x48, 0x49}.}
		\label{message}
	\end{figure*}
	It should be noted that the overall performance and throughput for any covert channel highly depends on the CPU's workload at the time of the communication. While executing numerous processes, it is likely for the context to be switched to an irrelevant process by the CPU, which can result in either a significant delay or invalidation of RSB entries.
	In our implementation, it can be shown that during each context switch, more than a single bit can be transferred to the receiver. Therefore, to achieve the optimal configuration, we analyzed the covert channel's performance through various settings. Table \ref{tab:covertchannel} demonstrates the measured bandwidths in our experiments, regarding the number of bits sent in each context switch. In our experiments, we reached the maximum bandwidth of 94.19 (KB/sec), by sending 3-bits per context switch, which required eight functions.   
	It can be interpreted from Table \ref{tab:covertchannel} that by increasing the number of transferred bits per context switch, the number of \textit{Flush} and \mbox{\textit{memory access}} operations, as well as the required memory, increases exponentially. Therefore, the overall delay caused by executing these operations would degrade the performance.  
	A significant amount of noise has been experienced in the previous covert channels such that the sender required error correction techniques such as cyclic redundancy check (CRC) to make the covert channels practical. In our proposed covert channel, the error rate is equal to 8.6E-5\%, which is considerably lower than previous work and therefore, no error correction technique is required.  
	Our covert channel increases the bandwidth compared to~\cite{tches-2021-30800}, by 3.76$\times$. Moreover, \cite{tches-2021-30800} has a 15\% to 25\% error rate. Our method can send more bits per context switch by increasing the number of gadgets. For instance, with four gadgets, in one context switch, the sender can transmit 00/01/10/11 to the receiver, and this increases the bandwidth, while \cite{tches-2021-30800} sends one bit for each context switch. Furthermore, the receiver in \cite{tches-2021-30800} should call \emph{N} nested functions in a \texttt{for} loop to determine the transmitted value, which reduces the bandwidth and increases the error rate. Furthermore, unlike the cache-based covert channel, our covert channel on RSB does not rely on assumptions such as memory deduplication. Also, our method is resilient against some cache-based covert channel countermeasures such as cache partitioning.
	Fig. \ref{message} demonstrates the transferring of a message (the word "\textit{HI}") by the mechanism explained earlier.


	\section{Countermeasures}
	\label{sec:countermeasures}
	
	In this section, we discuss various countermeasures against the attacks discussed and implemented in this work. We note that these countermeasures do not guarantee the complete mitigation of underlying vulnerabilities.
	

	\subsection{Mitigating \texttt{clflush}} 
	In order to incapacitate the attacker's ability to monitor the victim's memory accesses, the current implementation of the \texttt{clflush} instruction should be modified. As briefly mentioned and addressed by Yarom \textit{et al}. \cite{FR:8}, and Gruss \textit{et al.} \cite{FF:67}, current implementation of the \texttt{clflush} instruction poses an imminent threat to processors security and sets the background for various attacks.
	By analyzing our results, we conclude that ARM processors, which do not provide an unprivileged cache eviction instruction, have been less prone against side-channel attacks, forcing an attacker to face substantial difficulties to monitor victim's memory access without such instruction. Therefore, as hardware-level mitigation against our covert channel, we propose that similar to ARM processors, the \texttt{clflush} instruction should only be available in privileged mode on Intel processors. 
	
	\subsection{Mitigating the PMU} 
	We observe that obtaining a high-resolution time measurement plays a critical role in many well-known side-channel attacks. As discussed in Section \ref{sec:building blocks}, an attacker can achieve this goal in ARM processors, by exploiting the PMU. Therefore, introducing random noise to the PMU time measurements, while not in any capacity contradicts its initial design motives, will effectively prevent its exploitation.
	The attacker distinguishes a cache hit and a cache miss by the time variance of up to 12 execution cycles. While introducing hardware-level noise to the PMU will spoil the attacker's time measurements, the substantial performance degradation is still in the order of few execution cycles and is insignificant in the process and the application-level.

	\subsection{RSB Patches}
	RSB \textit{refilling} (also known as RSB \textit{stuffing}) have so far been the most promising software patches to mitigate underlying microarchitectural design flaws in RSB. However, this technique still cannot mitigate RSB vulnerabilities in some scenarios. 
	RSB refilling invalidates RSB entries at the time of a context switch, by inserting the address of a \textit{benign delay gadget} into RSB. Although effective in many cases, RSB refilling fails in case of an underfill, if the processor's microarchitectural design incorporates switching to BTB technique, discussed in Section \ref{subsec:RSB}. 
	We conclude that RSB must be flushed at every context switch, and switching to BTB should be avoided in case of an underfill. Our proposed solution does not suffer the performance degradation due to speculative execution of a benign gadget, as proposed by RSB refilling and mitigates threat models in which the attacker deliberately underfills RSB in order to exploit the BTB. 
	
	\section{Conclusion}
	\label{sec:conclustion}
	
	In this work, we investigated Spectre-family attacks on various ARM processors. To this end, we introduced and analyzed a number of critical building blocks to implement several experiments and attack scenarios. 
	We further introduced a high throughput and noise-free covert channel by exploiting RSB. The throughput of the proposed covert channel can reach to 94.19KB/s. 
	Finally, we discuss some countermeasures that can mitigate these attacks. We hope this work can cast light on previously neglected vulnerabilities, provide a better evaluation of common processors against Spectre-family security vulnerabilities and help reduce such threats in uprising areas such as IoT.
	We hope that this work effectively fill the gap for security analysis of previously neglected processors in various threat models and casts light on hidden and realistic vulnerabilities in commonly used processors, both in Intel and ARM family.

	
	%



	\ifCLASSOPTIONcompsoc


	\bibliography{ms.bib}
	\bibliographystyle{ieeetr}

	\begin{IEEEbiography}[{\includegraphics[width=1in,height=1.25in,clip,keepaspectratio]{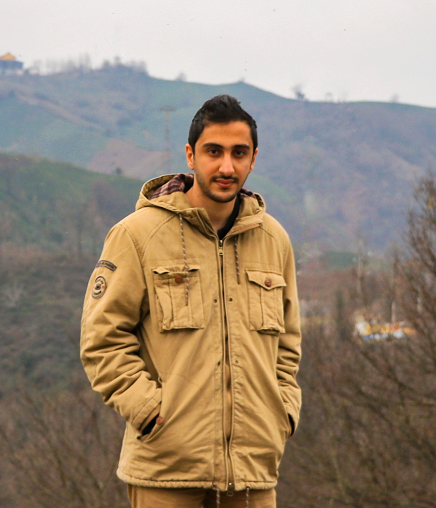}}]{Farhad Taheri Ardakani}
		received the B.S. degree in computer engineering from Shahid Bahonar University of Kerman (SBUK),
		Kerman, Iran, in 2016, and M.Sc. degree in computer engineering from Sharif University of Technology (SUT) in 2018.
		From 2016 to 2018, he is a member of the Data Storage, Networks, and Processing (DSN) Lab
		at SUT where he researched on reliability of Solid-State Drives (SSDs).
		Currently he is a Ph.D. student at the Smart and Secure Systems (3S) lab at SUT under supervision of Dr. S. Bayat Sarmadi.
		His research interests includes Privacy-Preserving Machine Learning, Multi-Party Computation, Computer Architecture, 
		and System Security.
	\end{IEEEbiography}
	
	\begin{IEEEbiography}[{\includegraphics[width=1in,height=1.25in,clip,keepaspectratio]{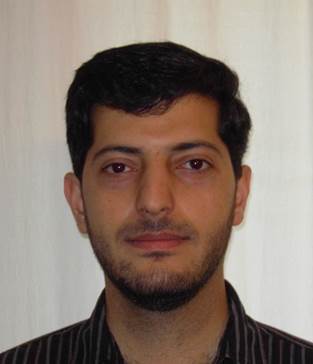}}]{Siavosh Bayat Sarmadi}
	received the B.Sc. degree from the University of Tehran, Iran, in 2000, the M.Sc. degree from Sharif University of Technology, Tehran, Iran, in 2002, and the PhD degree from the University of Waterloo in 2007, all in computer engineering (hardware). He was with Advanced Micro Devices, Inc. for about 6 years. Since September 2013, he has been a faculty member in the Department of Computer Engineering, Sharif University of Technology. He has served on the executive committees of several conferences. His research interests include hardware security and trust, cryptographic computations, and secure, efficient and dependable computing and architectures. He is a member of the IEEE.
	
	\end{IEEEbiography}

	\begin{IEEEbiography}[{\includegraphics[width=1in,height=1.25in,clip,keepaspectratio]{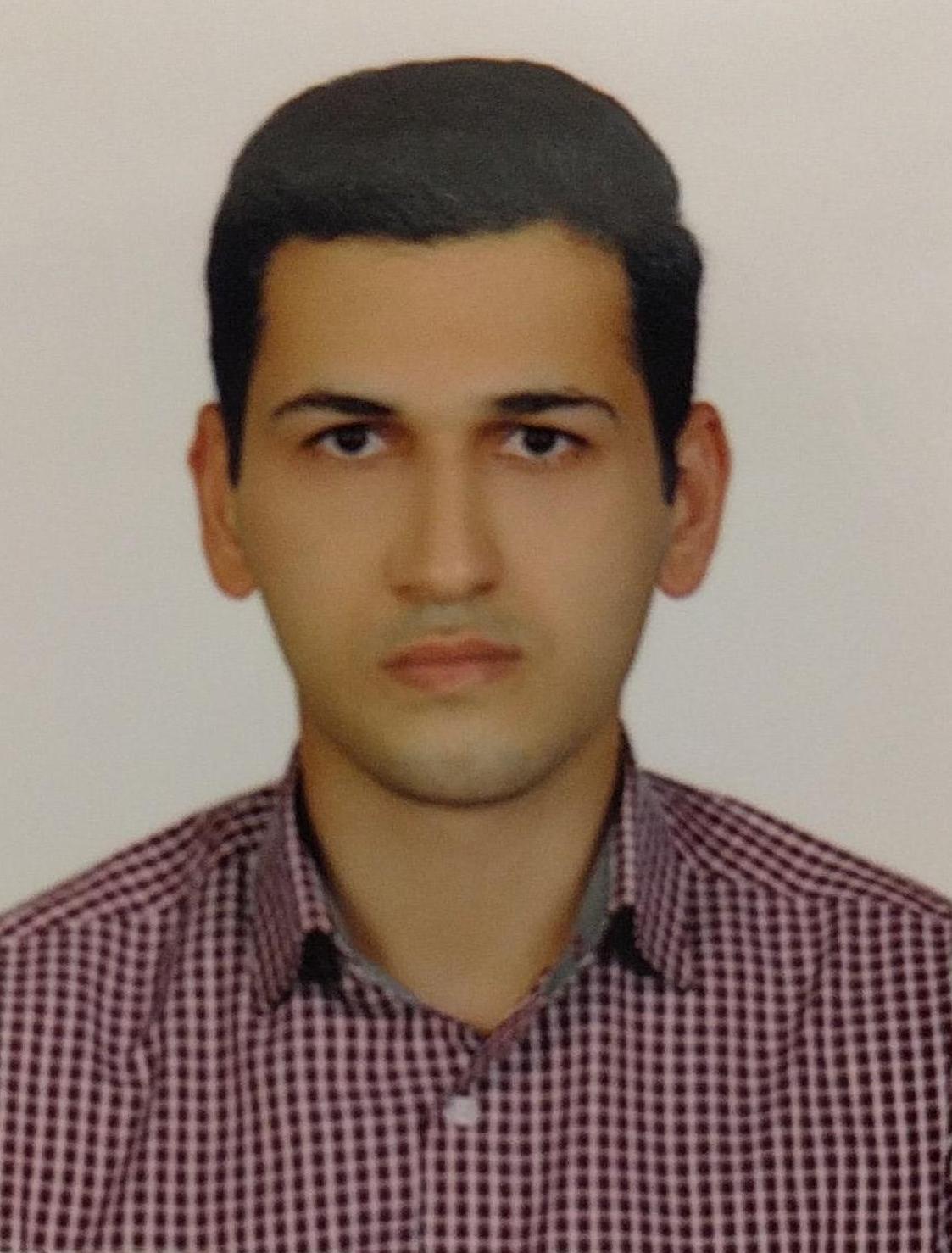}}]{Alireza Sadeghpour}
	 received the B.S. degree in computer engineering from Shahed University, Tehran, Iran, in 2018 and M.Sc. degree at 3S Lab in SUT under the supervision of Dr. Siavash Bayat Sarmadi. His research interests include Hardware security and Side-Channel Attack.

\end{IEEEbiography}

	\begin{IEEEbiography}[{\includegraphics[width=1in,height=1.25in,clip,keepaspectratio]{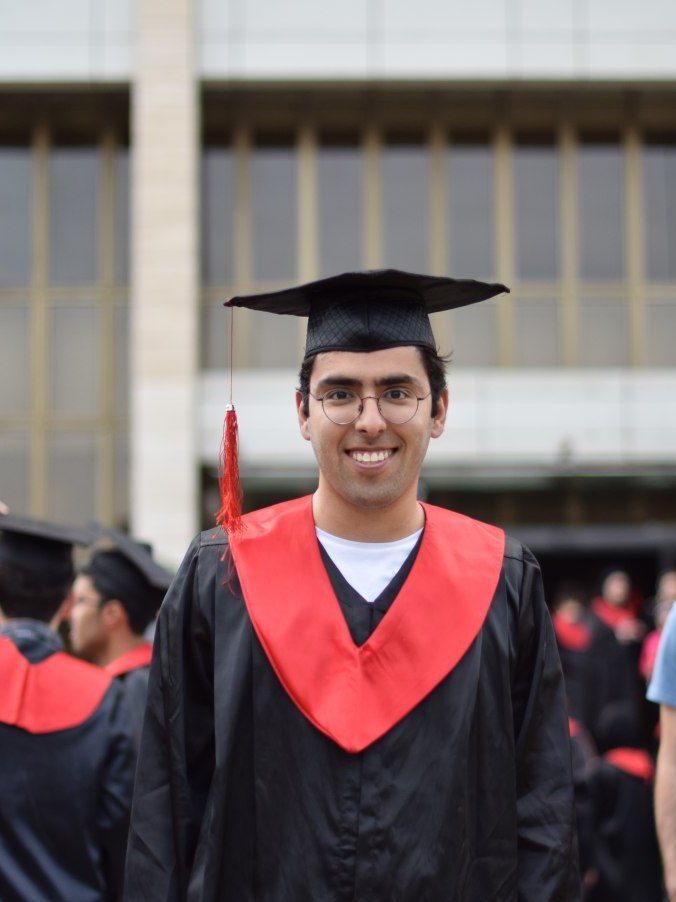}}]{Seyed Parsa Tayefeh Morsal}
	is in his last year of studies towards a B.Sc. in Computer Engineering at Sharif University of Technology. He is currently undertaking research in system security under the supervision of Dr. Bayat-Sarmadi in Smart and Secure Systems(3S) Laboratory. He is also a former lab member in S4Lab. His research interests include, but not limited to Side-channel attacks, Trusted computing, Hardware security, Vulnerability discovery \& Distributed systems.

\end{IEEEbiography}

\end{document}